\documentclass [12pt]{article}   
\usepackage{epsfig}
\usepackage{hangcaption}
\usepackage{array}
\title{Higher Dimensional Operator Corrections to the Goldstino
Goldberger-Treiman Vertices}

\author{Taekoon Lee\footnote{email: tlee@ctp.snu.ac.kr}
        \\
        \\
        Center for Theoretical Physics\\
        Seoul National  University\\
        Seoul 151-742,  Korea}

\date{}
\begin{document}
\maketitle
\begin{abstract}
The goldstino-matter interactions given by the Goldberger-Treiman relations can
receive higher dimensional operator corrections of $ O(q^{2}/M^{2})$, 
where $M$ denotes the mass of the mediators through which SUSY breaking 
is transmitted. These corrections in the gauge mediated SUSY breaking models
arise from loop diagrams, and an explicit calculation of such corrections is
presented. It is emphasized that the Goldberger-Treiman vertices are valid
only below the mediator scale and at higher energies goldstinos decouple
from the MSSM fields. The implication of this fact on gravitino cosmology in
GMSB models is mentioned.

\end{abstract}

\def\thepage{SNUTP-98-146}
\thispagestyle{myheadings}
\newpage
\pagenumbering{arabic}
\addtocounter{page}{0}
\newcommand{\be}{\begin{equation}}
\newcommand{\ee}{\end{equation}}
\newcommand{\bear}{\begin{eqnarray}}
\newcommand{\eear}{\end{eqnarray}}

The light gravitino to matter interactions are dominated by the spin $
\frac{1}{2}$ longitudinal component of gravitino which is essentially
the goldstino eaten by the gravitino via supersymmetric Higgs mechanism.
At energies much higher than the gravitino mass, the supersymmetric version
of the equivalence theorem \cite{fayet,equiv} allows one to replace the 
gravitino with the 
goldstino. The low energy interactions of a goldstino to matter fields,
which in this letter assumed to be the fields in the minimal supersymmetric
standard model (MSSM), are completely fixed model-independently by the so 
called goldstino Goldberger-Treiman vertices \cite{fayet,clark} which depend on 
the mass splittings of superparteners.
This is similar to the  Goldberger-Treiman relations in 
pion-nucleon interactions which also depend on the chiral symmetry breaking
parameters, namely, the nucleon masses.

At high energies the goldstino Goldberger-Treiman interactions are expected
to get corrections of $O(q^{2})$ where $q^{2}$ denotes generic Lorentz
invariants of the external momenta. At a first glance, one might think that
this correction is suppressed by $\frac{1}{F}$, where $F$ is the goldstino
decay constant, in analogy to the correction in  the pion-nucleon interaction
which is of $O(q^{2}/f_{\pi}^{2})$. However, unlike in the pion-nucleon 
case in which there is only one fundamental scale, namely,
$f_{\pi}$, there can be
multiple scales in realistic SUSY models, so it is possible that the
correction is suppressed by $\frac{1}{M_{X}^{2}}$, where $M_{X}$ is
an intermediate scale between the MSSM scale and $\sqrt{F}$.
If this is indeed the case, the corrections can be much larger than a
naive expectation based on the analogy to the Goldberger-Treiman relations
in hadron physics.
In this letter, we show that the corrections to the
goldstino Goldberger-Treiman couplings can be suppressed by an intermediate
scale by an explicit calculation of 
such corrections in gauge mediated SUSY breaking (GMSB)
models \cite{gm}.

In GMSB models, SUSY breaking occurs in a hidden sector and is transmitted to
the MSSM sector through gauge interactions between mediators and the
MSSM fields. The mediator scale in these models can be much lower than the
SUSY breaking scale. Because the SUSY breaking occurs in the hidden sector
there is no tree level coupling between the goldstino and the MSSM fields,
and the Goldberger-Treiman vertices are induced through loop diagrams.
Since the goldstino-matter interaction arise from loops, it becomes clear that
the correction to the the Goldberger-Treiman vertices can be
$O(q^{2}/M^{2})$, where $M$ denotes the mass of the mediators that go through
the loop diagrams, unless there is an exact cancellation among the 
diagrams. We shall see that such cancellation does not occur in GMSB models.

We consider a GMSB model in which there is a gauge singlet 
superfield $S$ through
which the hidden sector and the visible sector are connected. $S$ communicates
to the MSSM fields through interactions with  a set of mediators 
$\{q_{1i},q_{2i}\}$ via the superpotential
\be
{\cal L}_{w}=h \sum_{i}^{N_{q}} S \,q_{1i}\, q_{2i}
\label{e1}
\ee
where $N_{q}$ denotes the number of mediators, and $h$ is a coupling
constant. Note that $q_{1i},q_{2i}$
carry the opposite standard model gauge quantum numbers, respectively.
For our purpose, the details of the interaction between $S$ and the hidden
sector fields are not needed; The only requirement is that the vevs
$<\!F_{s}\!>$ and $<\!S\!>$ are nonvanishing.

For simplicity we first consider SUSY QED for the MSSM sector, since
the corrections to the Goldberger-Treiman vertices in nonabelian
gauge theories turns out to be identical as in the abelian case. 
The goldstino Goldberger-Treiman vertices between massless Weyl fermion
($\psi$), sfermion ($\phi$)
and gauge boson ($A_{\mu}$), gaugino ($\lambda$) are given by \cite{fayet,lw}
\be
{\cal L}_{\mbox{{\scriptsize GT}}} = \frac{m_{\phi}^{2}}{F} \chi \psi\phi^{*} +
\frac{im_{\lambda}}{\sqrt{2}F} \chi \sigma^{\mu\nu}\lambda F_{\mu\nu}
-\frac{ e m_{\lambda}}{\sqrt{2}F} \phi^{*}\phi\chi\lambda +h.c. \;\; ,
\label{e2}
\ee
where $\chi$ denotes goldstino, $m_{\phi}$, $m_{\lambda}$ are the
sfermion and gaugino masses, respectively, and $e$ is the gauge coupling.
Throughout this paper we follow the convention for spinors and metric given 
in \cite{bagger}, except that our gaugino $\lambda$ is related to the gaugino 
in \cite{bagger} by $\lambda=-i \lambda_{WB}$. Note that the metric in this 
convention is Diag(-1,1,1,1).

The interaction lagrangian in SUSY QED is given by
\bear
{\cal L}_{\mbox{{\scriptsize QED}}} &=& -e A_{\mu} \psi \sigma^{\mu} \overline{\psi}
+ i e A_{\mu} ( \phi^{*} \partial^{\mu} \phi -
\partial^{\mu} \phi^{*} \phi) \nonumber \\
                      && - \sqrt{2} e (  \phi^{*}\psi\lambda +
\phi\bar{\psi}\bar{\lambda}) -\frac{e^{2}}{2} (\phi^{*}\phi)^
{2} - e^{2} A_{\mu}A^{\mu}\phi^{*}\phi,
\label{e3}
\eear
and the couplings between the mediators and the SUSY QED fields are given by:
\bear
{\cal L}_{1} &=& -e A_{\mu}\left[ \Psi_{1i} \sigma^{\mu} \overline{\Psi}_{1i}
- \Psi_{2i} \sigma^{\mu} \overline{\Psi}_{2i}\right. + \nonumber \\
&& \left. i(\Phi_{1i}^{*} \partial^{\mu} \Phi_{1i} -\partial^{\mu} 
\Phi_{1i}^{*} \Phi_{1i}- \Phi_{2i}^{*} \partial^{\mu} \Phi_{2i} +
\partial^{\mu} \Phi_{2i}^{*} \Phi_{2i})\right] \nonumber \\
&& -\sqrt{2} e ( \Phi_{1i}^{*} \Psi_{1i}\lambda -
\Phi_{2i}^{*} \Psi_{2i}\lambda + h.c ) -e^{2}
\phi^{*} \phi ( \Phi_{1i}^{*} \Phi_{1i}-  \Phi_{2i}^{*} \Phi_{2i}),
\eear
where the last term arises from the $D$ term.

When SUSY is broken in the hidden sector and $F_{S}$ develops nonzero vev,
a mixing occurs between the spin half component $\psi_{S}$
of the chiral field $S$ and the goldstino from the hidden sector. Due to
the mixing, the true goldstino has a $\psi_{S}$ component given by
\be
\chi = -\frac{F_{S}}{F} \psi_{S} + \cdots
\ee
where the ignored terms involve only hidden sector fermions.
For small $F_{S}/F$, which is assumed in this letter, the above relation
can be inverted, giving
\be
\psi_{S} =\frac{F_{S}}{F} \chi + \cdots.
\ee
Now using the superpotential (\ref{e1}) we obtain the interaction between 
the goldstino and the mediators as
\be
{\cal L}_{2}= h \frac{F_{S}}{F}  (\Phi_{1i}\Psi_{2i}\chi +
\Phi_{2i}\Psi_{1i}\chi +h.c).
\ee
Note that this interaction is nothing but the Goldberger-Treiman vertex
in the mediator sector since $h F_{S}$ is the mass squared splitting
of the mediators.

From the above interactions, it is easy to see that the Goldberger-Treiman
vertices (\ref{e2}) arise from loop diagrams. The $\phi^{*}\psi\chi$
vertex in (\ref{e2}) comes from the two loop diagrams (Fig. 1) 
and the $A_{\mu}\lambda
\chi$, $\phi^{*}\phi \lambda\chi$ vertices arise from the one
loop diagrams in Fig.2 and Fig.3, respectively.
\begin{figure} 
\begin{center}
\epsfig{file=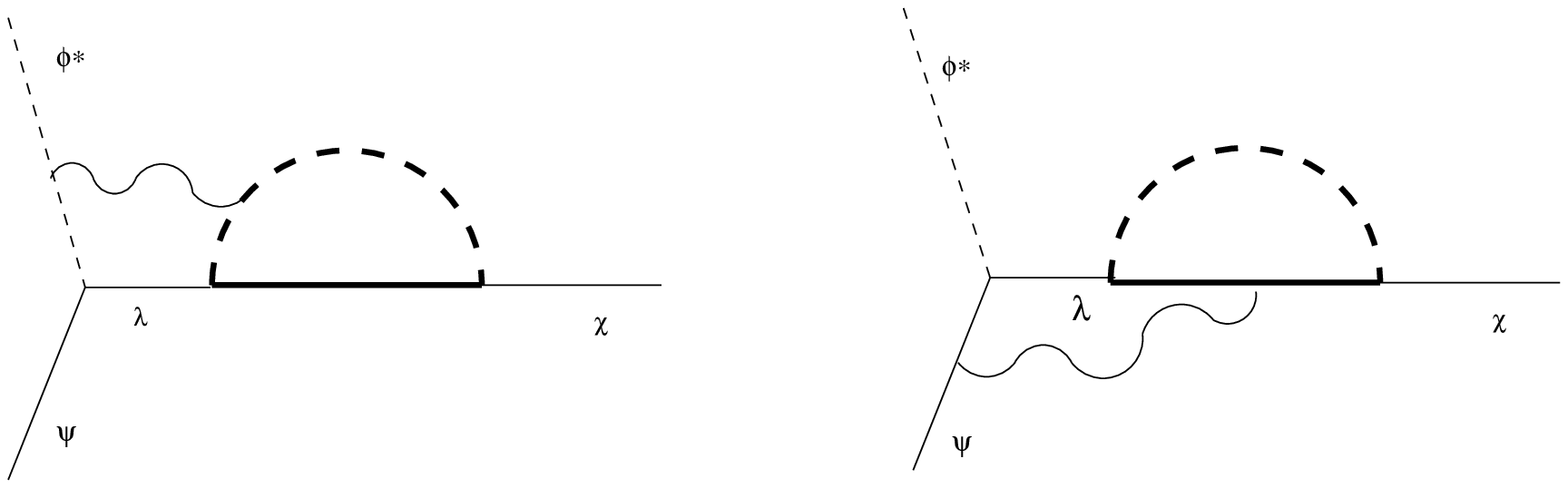, %
        height=10.0cm, angle=-90}
\end{center}
\isucaption{Examples of two loop diagrams that give rise to the
 $\phi^{*}\psi\chi$ Goldberger-Treiman vertex. Wavy lines denote
 gauge bosons and
 the thick solid and dashed lines denote fermionic and bosonic mediators,
respectively.}
\end{figure}
Phenomenologically, at high energies the dim-5 operators in
the Goldberger-Treiman vertices are more interesting
since the cross sections due to the dim-4 operator are always
suppressed by $O(m^{2}/s)$, where $m$ denotes the soft masses in MSSM 
and $s$ is the c.m. energy squared, compared to those from the dim-5 operators.
We therefore consider the higher dimensional 
operator corrections only for the dim-5 operators.

Let us first consider the $A_{\mu}\lambda \chi$ vertex in (\ref{e2}).
We first assume that the mass splitting  between the superparteners 
in the mediator sector is much smaller than the mediator mass. This 
requires 
\be
h S^{2} \gg F_{S}.
\ee
Then, as mentioned,  this vertex arises from the diagrams in Fig. 2.
There are other one loop diagrams; however, they are suppressed 
by $O(F_{S}/h S^{2})$ compared to those in Fig.2 and so can be ignored.
\begin{figure} 
\begin{center}
\epsfig{file=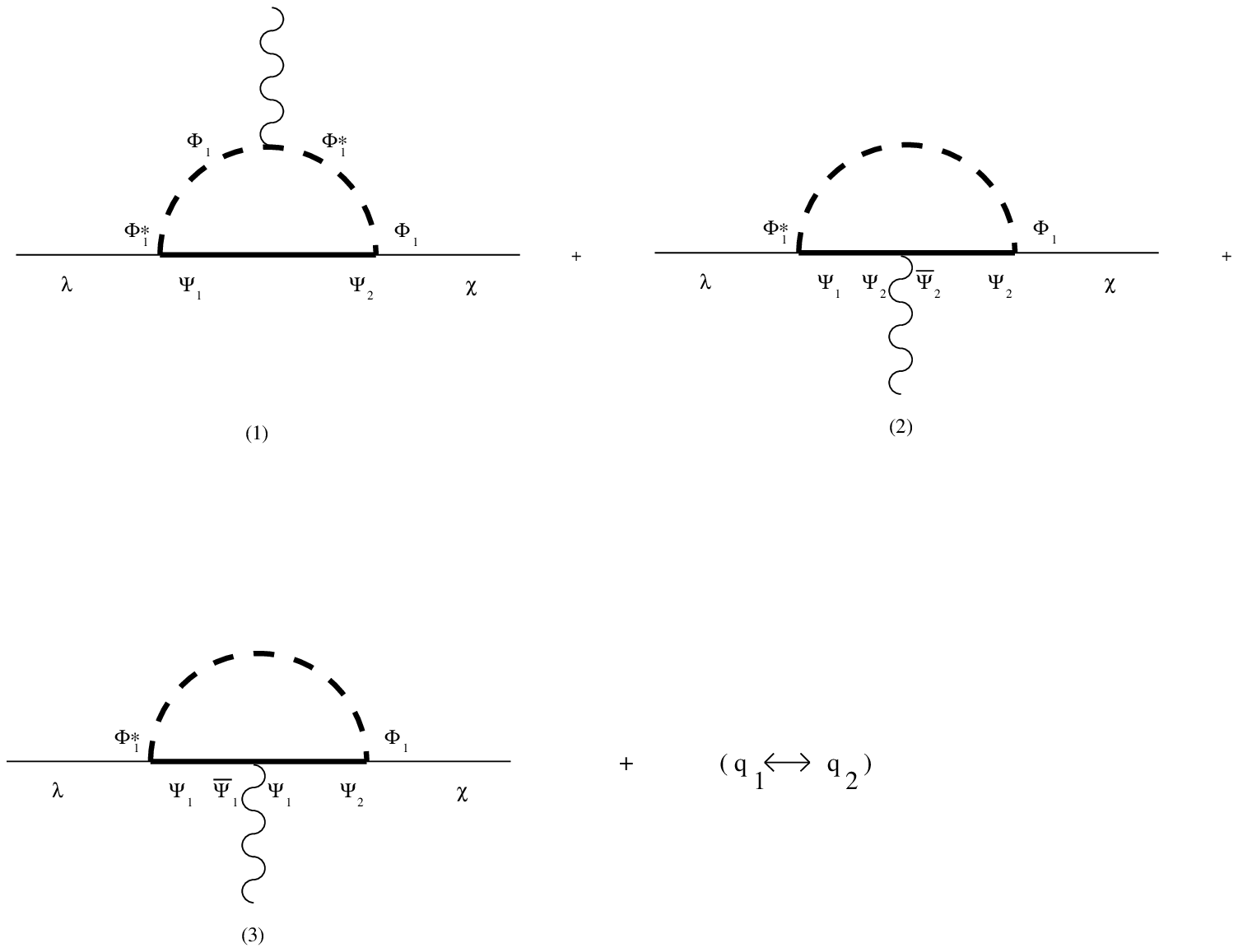, %
        height=8cm, angle=-90}
\end{center}
\isucaption{Diagrams that give rise to the
 $A_{\mu}\lambda\chi$ Goldberger-Treiman vertex. Wavy lines denote gauge bosons
 and the thick solid and dashed lines denote fermionic and bosonic mediators,
respectively. $q_{1}$ and $q_{2}$ denote mediators.}
\end{figure}
It is straightforward to calculate the diagrams. First the diagram (1) 
gives 
\bear
{\cal A}_{1}&=& i\frac{ \sqrt{2}h e^{2}N_{q} M F_{S}}{F}  
\int \prod_{i}^{3}d^{4} p_{i}
\tilde{\lambda}(p_{1})\tilde{\chi}(p_{2})\tilde{A}_{\mu}(p_{3})
(2\pi)^{4}\delta^{4}(\sum_{i}^{3}p_{i})I^{\left(1\right)}_{\mu}(p_{1},p_{2},M) 
\eear
where
\be
I^{\left(1\right)}_{\mu}(p_{1},p_{2},M) =\frac{i}{16\pi^{2}}\int\prod_{i}^{3}
d x_{i}\delta(\sum_{1}^{3} x_{i}-1)\left[\frac{
(1-2 x_{1})p_{1\mu}-(1-2x_{2})p_{2\mu}}{M^{2}-t^{2}+
x_{1} p_{1}^{2}+x_{2}p_{2}^{2}}\right]
\label{e10}
\ee
with 
\be
t_{\mu}=(x_{1} p_{1} -x_{2} p_{2})_{\mu}.
\ee
Here $x_{i}$ are the Feynman parameters, $M$ is the mediator mass
\be
M=h <\!S\!>,
\ee
and the Fourier transform is defined as
\be
\tilde{f}(p)=\int \frac{d^{4}p}{(2\pi)^{4}} f(x) e^{i p \cdot x}.
\ee

From the diagrams (2) and (3) we get
\bear
{\cal A}_{2}&=& -i \frac{\sqrt{2} h e^{2}N_{q} M F_{S}}{F}\int 
\prod_{i}^{3}d^{4} p_{i}
\tilde{\lambda}(p_{1})\sigma^{\mu}\bar{\sigma}^{\nu} \tilde{\chi}(p_{2})
\tilde{A}_{\mu}(p_{3})
(2\pi)^{4}\delta^{4}(\sum_{i}^{3}p_{i})\times \nonumber \\ 
&& \hspace{1.4in} I_{\nu}^{\left(2\right)}(p_{1},p_{2},M)
\eear
where
\be
I_{\nu}^{\left(2\right)}(p_{1},p_{2},M) =
-\frac{i}{16\pi^{2}}\int\prod_{i}^{3}d x_{i} 
\delta(\sum_{1}^{3}
x_{i}-1)\left[\frac{ (t+p_{2})_{\nu}}{M^{2}-t^{2}+x_{1} p_{1}^{2}+x_{2}
p_{2}^{2}}\right]
\label{e15}
\ee
and 
\bear
{\cal A}_{3}&=& -i\frac{ \sqrt{2} h e^{2}N_{q} M  F_{S}}{F}\int
\prod_{i}^{3}d^{4} p_{i}
\tilde{\lambda}(p_{1})\sigma^{\nu}\bar{\sigma}^{\mu} \tilde{\chi}(p_{2})
\tilde{A}_{\mu}(p_{3})
(2\pi)^{4}\delta^{4}(\sum_{i}^{3}p_{i}) \times \nonumber \\
&& \hspace{1.4in} I_{\nu}^{\left(3\right)}(p_{1},p_{2},M) 
\eear
with
\be
I_{\nu}^{\left(3\right)}(p_{1},p_{2},M) =-I_{\nu}^{\left(2\right)}
(p_{2},p_{1},M).
\ee

The three other diagrams obtained from diagrams (1), (2), and (3) by
exchanging the mediators $q_{1i} \leftrightarrow q_{2i}$ give identical
amplitudes to their corresponding diagrams. Now for small external
momenta compared to the mediator mass, we can expand in $1/M^{2}$
the denominators in $I^{\left(i\right)}$ and integrate over $x_{i}$ 
explicitly. Adding the six diagrams,
this gives to $O(p_{i}\cdot p_{j}/M^{2})$
\bear
{\cal A} &=&\sum_{1}^{6}{\cal A}_{i} \nonumber \\
&=&-\frac{\sqrt{2} m_{\lambda}}{F}\int
\prod_{i}^{3}d^{4} p_{i}
\tilde{\lambda}(p_{1})\sigma^{\mu\nu} 
\tilde{\chi}(p_{2}) \tilde{A}_{\mu}(p_{3}) (2\pi)^{4}
\delta^{4}(\sum_{i}^{3}p_{i})\times \nonumber \\
&& \left[ 1- \frac{1}{6 M^{2}}(p_{1}^{2}+p_{1}\cdot p_{2} +p_{2}^{2})\right]
(p_{1} +p_{2})_{\nu}
\eear
where $m_{\lambda}$ is the one-loop gaugino mass \cite{gm}
\be
m_{\lambda}=\frac{2 e^{2}N_{q} F_{S}}{16 \pi^{2} <\!S\!>}.
\ee
Converting this to the coordinates space we obtain the Goldberger-Treiman
vertex for $A_{\mu}\lambda\chi$ and its higher dimensional operator
correction:
\be
{\cal A} = \int d^{4} x {\cal L}_{\chi\lambda A_{\mu}}(x)
\ee
with
\bear
 {\cal L}_{\chi\lambda A_{\mu}} &=&
\frac{im_{\lambda}}{\sqrt{2}F}\left[
\chi \sigma^{\mu\nu}\left( 1+ \frac{1}{6 M^{2}}(
{\stackrel{\leftarrow}{\partial}}^{2} +\stackrel{\leftarrow}{
\partial} \cdot \stackrel{\rightarrow}{\partial} + \stackrel{\rightarrow}
{\partial}^{2})\right)
\lambda\right] F_{\mu\nu}.
\label{e21}
\eear
Note that no on-shell condition for the goldstino was used in deriving
(\ref{e21}).
\begin{figure} 
\begin{center}
\epsfig{file=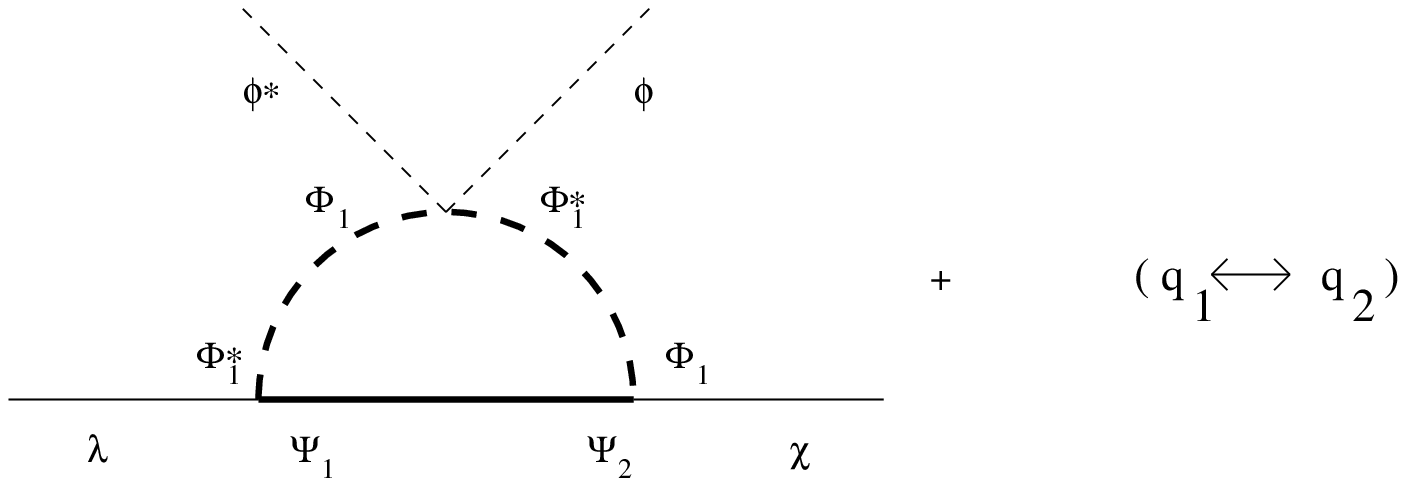, %
        height=9.0cm, angle=-90}
\end{center}
\isucaption{Diagrams that give rise to the
 $\chi\lambda\phi^*\phi$ Goldberger-Treiman vertex. 
 The thick solid and dashed lines denote
 fermionic and bosonic mediators,
respectively. $q_{1},q_2$ denote mediators.}
\end{figure}

The other dim-5 vertex of $\phi^{*}\phi\lambda\chi$ in the Goldberger-Treiman
interaction arises from the two diagrams in Fig.3. A straightforward
calculation of the diagrams gives:
\bear
 i\frac{2 \sqrt{2}h e^{3}N_{q} M F_{S}}{F} \int \prod_{i}^{3}d^{4} p_{i}
\tilde{\lambda}(p_{1})\tilde{\chi}(p_{2})\widetilde{\phi^{*}\phi}(p_{3})
(2\pi)^{4}\delta^{4}(\sum_{i}^{3}p_{i}) I(p_{1},p_{2},M)
\eear
where
\be
I(p_{1},p_{2},M) =\frac{i}{16\pi^{2}}\int\prod_{i}^{3} d x_{i}
\delta(\sum_{1}^{3}
x_{i}-1)\left[\frac{1}{M^{2}-t^{2}+x_{1} p_{1}^{2}+x_{2}
p_{2}^{2}}\right].
\label{e23}
\ee
Expanding in $1/M^{2}$ the denominator  of the integrand in $I$ and 
integrating over $x_{i}$
we obtain the Goldberger-Treiman vertex and its correction as:
\be
{\cal A}_{\phi^{*}\phi\lambda\chi}
= \int d^{4} x {\cal L}_{\phi^{*}\phi\lambda\chi}(x)
\ee
with
\bear
{\cal L}_{\phi^{*}\phi\lambda\chi} &=&
-\frac{e m_{\lambda}}{\sqrt{2}F}\left[
\chi\left( 1+ \frac{1}{6 M^{2}}(
{\stackrel{\leftarrow}{\partial}}^{2} +\stackrel{\leftarrow}{
\partial} \cdot \stackrel{\rightarrow}{\partial} + \stackrel{\rightarrow}
{\partial}^{2})\right)\lambda\right] \phi^{*}\phi.
\label{e25}
\eear
Note that the higher dimensional operator corrections in (\ref{e21}) and
(\ref{e25}) are independent of the number of the mediators $N_{q}$.

For nonabelian gauge theory, the corrections to the Goldberger-Treiman
vertices can be found in essentially the same way as in the abelian case.
For SUSY QCD, for example, the Goldberger-Treiman vertices including 
the higher dimensional operator corrections for the dim-5 operators
are given by
\bear
{\cal L}^{\mbox{{\scriptsize QCD}}}_{\mbox{{\scriptsize GT}}}  & = &
 \frac{m^{2}_{\phi} - m^{2}_{\psi}}{F} \chi \psi_{i} \phi^{*}_{i}
 + \frac{im_{\lambda}}{\sqrt{2}F}\left[
\chi \sigma^{\mu\nu}\left( 1+ \frac{1}{6 M^{2}}(
{\stackrel{\leftarrow}{\partial}}^{2} +\stackrel{\leftarrow}{
\partial} \cdot \stackrel{\rightarrow}{\partial} + \stackrel{\rightarrow}
{\partial}^{2})\right)\lambda^{a}\right] F^{a}_{\mu\nu} \nonumber \\
& & -\frac{ g m_{\lambda} }{ \sqrt{2} F} \left[
\chi\left( 1+ \frac{1}{6 M^{2}}(
\stackrel{\leftarrow}{\partial}^{2} +\stackrel{\leftarrow}{
\partial} \cdot \stackrel{\rightarrow}{\partial} + \stackrel{\rightarrow}
{\partial}^{2})\right)\lambda^{a}\right] 
\phi_{i}^{*}T^{a}_{ij} \phi_{j} \; + h.c. \;\; ,
\eear
where $g$ is the gauge coupling and $T^{a}_{ij}$ are the gauge
group generators.

\begin{figure} 
\begin{center}
\epsfig{file=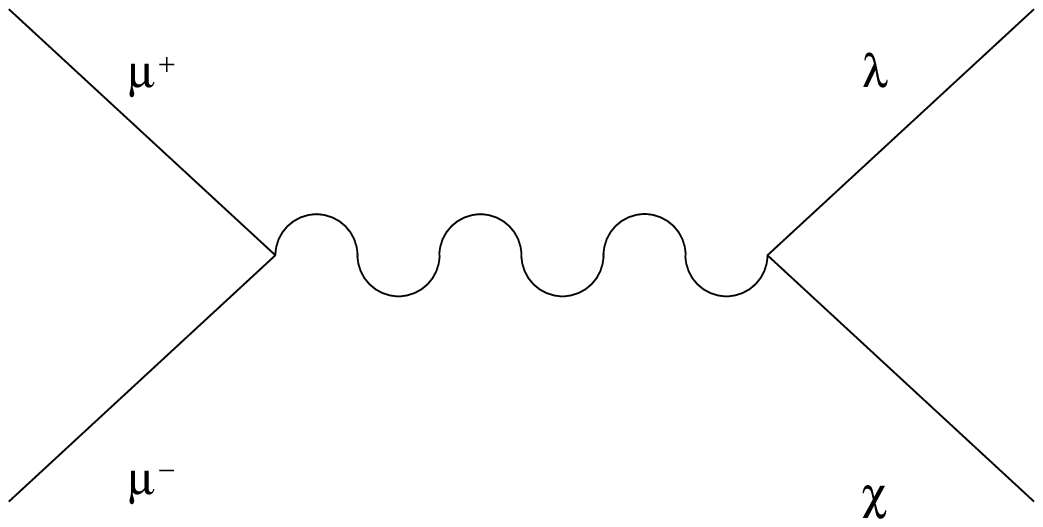, %
        height=5.0cm, angle=-90}
\end{center}
\isucaption{Diagram for $\mu^-\mu^+\rightarrow\chi\lambda$.}
\end{figure}
These higher dimensional operator corrections can affect the 
goldstino production rate at high energy scattering near the threshold 
of the mediator particles. Consider, for example,
$\mu^{-}\mu^{+} \rightarrow \chi\lambda$ process which were studied in Refs.
\cite{fayet2,dicus,nano}.
The dominant amplitude for the process comes from the diagram in Fig.4.
The cross section to $O(s/M^{2})$ from this diagram 
using the interaction  (\ref{e21}) is:
\be
\sigma=\sigma_{0}( 1+\frac{s}{6 M^{2}})
\ee
where
\be
\sigma_{0}=\frac{e^2 m_{\lambda}^{2}}{24\pi F^{2}}
\ee
and $s$ is the c.m. energy squared. This shows that, at $\sqrt{s}=M$, for
example, the goldstino production rate is increased about 17\% compared to
that obtained without the higher dimensional operator correction.
Of course, it would be very challenging to observe the direct production of
goldstinos in GMSB models since in these models $\sqrt{F}$ is
generally too large for accelerator access. However, in models in 
which the SUSY breaking scale is accessible to accelerators, 
the higher dimensional operator corrections to the Goldberger-Treiman vertices
could be used in probing the underlying SUSY breaking mechanism.

The corrections studied here can also have an important consequence in 
gravitino cosmology in GMSB models.
The fact that the goldstino-matter interaction arises from loop
diagrams indicates that goldstinos decouple from the MSSM fields
at energies above the mediator mass. This becomes clear from
Eqs. (\ref{e10}), (\ref{e15}) and (\ref{e23})
which show that at high energies the goldstino-matter
couplings decrease in proportion to $M^{2}/E^{2}$, where $E$
denotes the energy scale of the process in consideration, compared to the
Goldberger-Treiman vertices.
This also implies that in early universe light gravitinos decouple
linearly in $M^{2}/T^{2}$ from
the MSSM fields at temperature higher than the mediator mass.
It is therefore clear that the 
Goldberger-Treiman vertices are valid only below the mediator scale,
and cannot be used at energies higher than the mediator mass.
However, this fact has not been taken into account in the existing
bound on the reheating temperature obtained from
the gravitino overproduction \cite{moroi}. When the decoupling is taken into
account, one can expect that gravitino production at temperaturs above the
mediator scale is mostly due to the mediators whereas the MSSM fields 
contribution to the gravitino production is highly suppressed. 
This issue is currently under investigation \cite{tl}.

\vspace{.3in}
\noindent
{\bf Acknowledgements:}
The author is grateful to  K. Choi and H.B. Kim for useful
conversation, and owes special thanks to Guo-Hong Wu for useful discussions
at the initial stage of the work. This work was supported in part by the Korean
Science and Engineering Foundation.

 \end{document}